\documentclass[a4paper,twoside,reqno]{bjp}

\usepackage{graphicx}
\usepackage{cite}
\usepackage{amssymb,amsmath,amscd,amsthm}
\usepackage{times}

\usepackage{geometry}
\allowdisplaybreaks

\newcommand\tit\textit

\newcommand\gGL{\text{\normalfont GL}}

\newcommand\gSp{\text{\normalfont Sp}}

\newcommand\gO{\text{\normalfont O}}
\newcommand\gSO{\text{\normalfont SO}}
\newcommand\gPin{\text{\normalfont Pin}}
\newcommand\gSpin{\text{\normalfont Spin}}

\newcommand\aA{\mathfrak{A}}
\newcommand\aCl{\text{Cl}} 
\newcommand\lh{\mathfrak{h}}

\newcommand\lgl{\mathfrak{gl}}

\newcommand\lsu{\mathfrak{su}}
\newcommand\lsp{\mathfrak{sp}}
\newcommand\lo{\mathfrak{o}}

\begin{document}

\title{Fock space dualities}

\runningheads{Fock space dualities}{K. Neerg\aa rd}

\begin{start}{

\author{K. Neerg\aa rd}{}

\index{Neerg\aa rd, K.}

\address{N\ae stved, Denmark}{}

\received{27 August 2021}

}

\begin{Abstract}
Several cases of Fock space duality occurring in the theory of
many-body systems in general and nuclei in particular are discussed.
All of them are special cases of a general duality theorem proved in
mathematics by Howe in the 1970s. Dualities on a fermion Fock space
between orthogonal Lie algebras and related groups, including an
$\lo$--$\gPin$ duality recently discovered by the author, present a
nice, symmetric pattern.
\end{Abstract}

\begin{KEY}
Fock space, classical groups, duality.
\end{KEY}

\end{start}

\section{Introduction\label{sec:in}}

Fock space dualities first appeared in physics in the context of the
exploration of the new quantum mechanics in the late 1920s. Wigner and
von Neumann showed~\cite{ref:Wig28} that, by total antisymmetry, the
wave function
\begin{equation}
  \phi((m_{l1},m_{s1}) , \dots , (m_{ln_\text{el}},m_{sn_\text{el}}))
\end{equation}
of $n_\text{el}$ electrons in an atomic shell with azimutal quantum
number $l$, where $m_l$ and $m_s$ are the orbital and spin magnetic
quantum numbers, can be expanded on terms of the form
\begin{equation}\label{eq:WvN}
  \sum_\nu \chi^\lambda_\nu (m_{l1} , \dots , m_{ln_\text{el}})
      \psi^\lambda_\nu (m_{s1} , \dots , m_{sn_\text{el}}) .
\end{equation}
Here the functions $\chi^\lambda_\nu$ and $\psi^\lambda_\nu$ form
conjugate bases for conjugate irreducible representations (irreps) of
the groups of permutations of their arguments. That is, the matrix
representing a given permutation $s$ in one irrep and the inverse
transpose of that representing it in the other one are identical when
$s$ is even and differ by a factor $-1$ when $s$ is odd. The
equivalence classes of these irreps are described by conjugate Young
diagrams $\lambda$ and $\tilde\lambda$ like the following pair, where
$\lambda=(2,2,1,1,1)$ in terms of row lengths.

{\centering $\lambda=$
\raisebox{-14bp}{\includegraphics{young1}}\qquad $\tilde\lambda=$
\raisebox{-4bp}{\includegraphics{young2}}
\par}

The Young diagram $\lambda$ can have no more than $2l+1$ rows and no
more than 2 columns. Its area equals $n_\text{el}$. By the
correspondence between symmetry and equivalence class of irrep of the
general linear group $\gGL(n)$ of non-singular linear transformations
of an $n$-dimensional vector space over the complex numbers, or
equivalently, its Lie algebra $\lgl(n)$, the functions~\eqref{eq:WvN}
carry products of irreps of $\lgl(2l+1)$, acting on the arguments
$m_l$, and $\lgl(2)$, acting on the arguments $m_s$, and the
equivalence classes of these irreps are described by the Young
diagrams $\lambda$ and $\tilde\lambda$.

The atomic shell is a Fock space. The general Fock space is associated
with $k$ kinds of particle, where each kind may be a kind of fermion
or a kind of boson. The particles inhabit a common single-kind state
space of dimension $d$. I label the particle kinds by letters
$\tau, \upsilon, \dots$ and basic single-kind states by letters
$p, q, \dots$\,. The creation operator of a particle of kind $\tau$ in
the state $| p \rangle$ is denoted by $a^\dagger_{p\tau}$ and the
corresponding annihilation operator by $a_{p\tau}$. These operators
obey the usual commutation relations, but contrary to quantum
mechanical conventions, $a^\dagger_{p\tau}$ and $a_{p\tau}$ are
\emph{not assumed Hermitian conjugate}. I impose, indeed, no Hermitian
inner product on any state space. The \tit{Fock space} of the system
thus described is spanned by the states generated from the vacuum by
the operators
\begin{equation}
  1 , \quad a^\dagger_{p\tau} , \quad
  a^\dagger_{p\tau} a^\dagger_{q\upsilon} , \quad \dots
\end{equation}
It has finite dimension in the case of only fermions and infinite
dimension in the presence of bosons. In the atomic example, electrons
with spins up and down are viewed as different kinds and the
single-kind state space is spanned by the states $| m_l \rangle$. The
following is then an alternative formulation of the observations
above.

\tit{$\lgl(2l+1)$--$\lgl(2)$ duality: The Fock space of the atomic
shell has the decomposition
\begin{equation}\label{eq:atom}
   \Phi = \bigoplus \textup{X}_\lambda \otimes \Psi_\lambda ,
\end{equation}
where the sum runs over all Young diagrams $\lambda$ with at most
$2l+1$ rows and at most 2 columns and $\textup{X}_\lambda$ and
$\Psi_\lambda$ carry irreps of $\lgl(2l+1)$ and $\lgl(2)$ described by
the Young diagrams $\lambda$ and $\tilde\lambda$ and acting on the
variables $m_l$ and $m_s$, respectively.}

Similar decompositions apply to many different Fock spaces, some of
which I describe in this paper. A structure like~\eqref{eq:atom} is
called a \tit{duality} in the literature.

\section{Symplectic and orthogonal groups and Lie algebras and their
number conserving and number non-conserving realisations on Fock
spaces\label{sec:symor}}

The symplectic group $\gSp(n)$ and orthogonal group $\gO(n)$ are
subgroups of $\gGL(n)$ defined by non-degenerate bilinear forms $b$.
In terms of basic vectors $| i \rangle$ for the defining vector space
of $\gGL(n)$, their members $g$ obey
\begin{equation}
  \sum_{kl} \langle b | k l \rangle \langle k | g | i \rangle
      \langle l | g | j \rangle = \langle b | i j \rangle .
\end{equation}
When $b$ is \tit{skew symmetric}, the group is \tit{symplectic}, when
$b$ is \tit{symmetric}, it is \tit{orthogonal}. For a given $n$, all
symplectic groups are isomorphic and all orthogonal groups are
isomorphic, independently of $b$. I denote the Lie algebras of
$\gSp(n)$ and $\gO(n)$ by $\lsp(n)$ and $\lo(n)$. The group $\gSp(n)$
only exists for even $n$, the group $\gO(n)$ for every $n$. While
$\gSp(n)$ is simply connected, $\gO(n)$ is not even connected and its
maximal connected subgroup $\gSO(n)$ of index 2 formed by the
orthogonal transformations with determinant 1 not simply connected.
(The transformations in the coset have determinant $-1$.) Both
$\gO(n)$ and $\gSO(n)$ have double covering groups $\gPin(n)$ and
$\gSpin(n)$, where the latter is simply connected and a subgroup of
the former of index 2. The equivalence classes of irreps of $\lsp(n)$
and $\lo(n)$ correspond 1--1 to those of the simply connected groups
$\gSp(n)$ and $\gSpin(n)$. The \tit{spin irreps} of $\gSpin(n)$ are
double valued on $\gSO(n)$ and likewise those of $\gPin(n)$ on
$\gO(n)$. The equivalence classes of $\lsp(n)$ and $\lo(d)$ irreps are
described by Young diagrams with at most $n/2$ rows. The $\lsp(n)$
Young diagrams are ordinary ones~\cite{ref:Wey39}. To describe spin
$\lo(n)$ irreps (which expand to spin irreps of $\gSpin(n)$) one also
needs Young diagrams with half-integral row lengths like the following
with row lengths $9/5,7/2,3/2$.

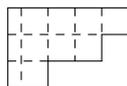
{\centering\includegraphics{young6}\par}

Further, for even $n$, the Young diagrams of maximal depth come in
pairs with opposite signs of the ``lengths'' of their bottom rows like
the following pair for $n=6$ with row lengths $4,3,\pm2$.

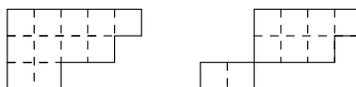
{\centering\includegraphics{young28}\qquad
\centering\includegraphics{young29}\par}

I call \tit{mirrors} the associated equivalence classes of $\lo(d)$
irreps. For $n=2$ the edge whence the single row extends must be
specified. (The abelian Lie algebra $\lo(2)$ actually has a continuum
of inequivalent 1-dimensional irreps. Those described by the present
Young diagrams are the only ones that occur in the theorems
below.)~\cite{ref:Nee20a}

The single-kind state space may be taken as defining vector space for
$\lsp(d)$ and $\lo(d)$. These Lie algebras then have realisations on
any Fock space given by
\begin{equation}\label{eq:nc}
  g \mapsto \sum_{pq\tau} \langle p | g | q \rangle
    a^\dagger_{p\tau} a_{p\tau}
\end{equation}
for an arbitrary member $g$ of the Lie algebra. These operators are
seen to conserve the number of particles. I reserve the symbols
$\lsp(d)$ and $\lo(d)$ for these realisations, which I call
\emph{number conserving}. For systems of only fermion or only bosons,
the Lie algebras $\lsp(2k)$ and $\lo(2k)$ have realisations spanned by
the operators
\begin{equation}\label{eq:nnc}
  \sum_p a^\dagger_{p\tau} a_{p\upsilon} \mp \frac d 2 ,  \qquad
  \sum_{pq} \, \langle p q | b \rangle
    a^\dagger_{p\tau} a^\dagger_{q\upsilon} , \qquad
  \sum_{pq} \, \langle b | p q \rangle a_{p\tau} a_{q\upsilon}
\end{equation}
with $-$ for fermions and $+$ for bosons. Here the matrix element
$\langle p q | b \rangle$ of the dual bilinear form is defined by
\begin{equation}
  \sum_r \langle b | p r \rangle \langle q r | b \rangle
  = \delta_{pq} .
\end{equation}
The operators in the last two sets in \eqref{eq:nnc} evidently do not
conserve particle number. (The sets are empty for $k=1$ in the cases
of fermions and $\lo(2k)$ and bosons and $\lsp(2k)$.) I reserve the
symbols $\lsp(2k)$ and $\lo(2k)$ for these realisations, which I call
\emph{number non-conserving}. The sets of operators \eqref{eq:nc} and
\eqref{eq:nnc} commute.

\section{Helmers's theorem and its orthogonal analogon\label{sec:Hel}}

In 1961, in the wake of the nuclear BCS theory, Helmers proved the
following~\cite{ref:Hel61}.

\tit{$\lsp(d)$--$\lsp(2k)$ duality (Helmers): For even $d$, a fermion
Fock space has the decomposition
\begin{equation}\label{eq:sp-sp}
   \Phi = \bigoplus \textup{X}_\lambda \otimes \Psi_\mu ,
\end{equation}
where $\lsp(d)$ acts on $\textup{X}_\lambda$ producing an irrep with
the Young diagram $\lambda$ and $\lsp(2k)$ acts on $\Psi_\mu$
producing an irrep with the Young diagram $\mu$. The sum runs over all
such pairs $(\lambda,\mu)$ that $\lambda$ and a rotated and reflected
copy of $\mu$ fill a $k \times d/2$ rectangle without overlap as in
the following example for $d=12$ and $k=4$, where
$\lambda = (4,3,2,2,1,1)$ and $\mu=(5,4,2)$.}

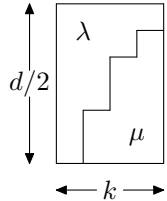
{\centering\includegraphics{young27}\par}

Much later, in 2019, i obtained the following orthogonal analogon of
Helmers's theorem~\cite{ref:Nee19}.

\tit{$\lo(d)$--$\lo(2k)$ duality: A fermion Fock space has the
decomposition
\begin{equation}\label{eq:o-o}
   \Phi = \bigoplus \textup{X}_\lambda \otimes \Psi_\mu ,
\end{equation}
where $\lo(d)$ acts on $\textup{X}_\lambda$ producing a representation
associated with the Young diagram $\lambda$ and $\lo(2k)$ acts on
$\Psi_\mu$ producing a representation associated with the Young
diagram $\mu$. The sum runs over all such pairs $(\lambda,\mu)$ that
$\lambda$ and a rotated and reflected copy of $\mu$ fill a
$k \times d/2$ rectangle without overlap as in the following example
for $d=11$ and $k=4$, where $\lambda = (4,3,2,2,1)$ and
$\mu=(9/2,7/2,3/2,1/2)$.}

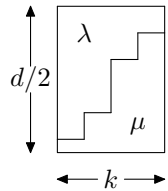
{\centering\includegraphics{young30}\par}

\tit{If the border between $\lambda$ and the copy of $\mu$ hits the
left edge of the rectangle (as in the example), the $\lo(d)$
representation is irreducible while the $\lo(2k)$ representation is
the direct sum of mirror irreps. If the border hits the bottom edge
(which requires that $d$ is even), the $\lo(2k)$ representation is
irreducible while the $\lo(d)$ representation is the direct sum of
mirror irreps.}

The proofs of both theorems are based on comparison of the characters
on both sides of the equations~\eqref{eq:sp-sp} and \eqref{eq:o-o}.

Several special cases of these theorems have applications in nuclear
physics. Thus for $k=1$, the number non-conserving Lie algebra
$\lsp(2k)$ is closely related to Kerman's \tit{quasispin}
algebra~\cite{ref:Ker61}, usually described as an $\lsu(2)$ algebra.
The latter is a Lie algebra over the reals whose complexification is
isomorphic to $\lsp(2)$. The $\lsp(d)$--$\lsp(2)$ duality connects the
quantum numbers of quasispin and \tit{seniority} as exploited
extensive in the work of Talmi, in particular~\cite{ref:Tal93}. For
$k=2$, the Lie algebra $\lsp(2k)=\lsp(4)$ is identical to the Lie
algebra $\lo(5)$ introduced by Flowers and Szpikowski to describe
systems with both neutrons and protons~\cite{ref:Flo64a}. For $k=4$,
corresponding to the 4-dimensional space of the nucleonic spin and
isospin, $\lo(2k)$ is the Lie algebra $\lo(8)$ advanced by these
authors as a ``quasispin algebra for $LS$
coupling''~\cite{ref:Flo64b}. Both Lie algebras $\lo(5)$ and $\lo(8)$
attracted attention around the turn of millennium in discussions of
pairing in nuclei including isospin $T=0$ paring.

\section{Howe's theorem\label{sec:Howe}}

All dualities mentioned so far are special cases of a very general
theorem proved in mathematics by Howe in the 1970s~\cite{ref:How89}.
Presenting his theorem requires more definitions. The following
formalism is not that of Howe but a ``physicists version'' based on
second quantisation~\cite{ref:Nee20a}. The Fock space is the general
one with possibly both fermion and bosons. Let $G$ be a ``classical''
group~\cite{ref:Wey39}, $\gGL(d)$, $\gSp(d)$ or $\gO(d)$, with the
single-kind state space as the defining vector space. It has a
realisation on the Fock space, which I denote by the same symbol as
the abstact group. I denote by $g$ an arbitrary member of $G$ and by
$\gamma$ its realisation on the Fock space. For a given $\tau$, the
realisation may be either \emph{cogredient}, that is,
\begin{equation}
  \gamma a^\dagger_{p\tau} = \left( \sum_q \langle q | g | p \rangle
    a^\dagger_{q\tau} \right) \gamma, \quad
  \gamma a_{p\tau} = \left( \sum_q \langle p | g^{-1} | q \rangle
    a_{q\tau} \right) \gamma .
\end{equation}
for every $g$, or \emph{contragredient}, that is,
\begin{equation}
  \gamma a^\dagger_{p\tau} = \left( \sum_q \langle p | g^{-1} | q \rangle
    a^\dagger_{q\tau} \right) \gamma , \quad
  \gamma a_{p\tau} = \left( \sum_q \langle q | g | p \rangle
    a_{q\tau} \right) \gamma .
\end{equation}
for every $g$. In the cases of $\gSp(d)$ and $\gO(d)$, co- and
contragredient actions are equivalent because the bilinear form $b$
provides a similarity transformations between the matrices
$\langle p | g | q \rangle$ and $\langle q | g^{-1} | p \rangle$.

Acting on the operators $a^\dagger_{p\tau}$ and $a_{p\tau}$, I define
a bracket $|\cdot,\cdot|$ which equals the anticommutator
$\{\cdot,\cdot\}$ when both operands are fermion operators, otherwise
the commutator $[\cdot,\cdot]$, so that their commutation relations
can be written
\begin{equation}
  |a^\dagger_{p\tau},a^\dagger_{q\upsilon}|
  = |a_{p\tau},a_{q\upsilon}| = 0 , \quad
  |a_{p\tau},a^\dagger_{q\upsilon}| = \delta_{p\tau,q\upsilon} .
\end{equation}
I define another bracket $]\cdot,\cdot[$ which is the complete
opposite, being equal to the commutator when both operands are fermion
operators and otherwise the anticommutator. Both brackets can be
extended to the span $\aA$ of the set of creation and annihilation
operators. The bracket $|\cdot,\cdot|$ can be extended further to the
set
\begin{equation}
  \bar \lh = \text{span} \, \{ \,a b \; | \;
     a,b \in \aA \, \} ,
\end{equation}
which, by the commutation relations, includes the numbers. By
definition, $|ab,cd|=[ab,cd]$ when either both $a$ and $b$ or both $c$
and $d$ are fermion operators or both of them are boson operators, and
$|ab,cd|=\{ab,cd\}$ when both $ab$ and $cd$ are products of one boson
operator and one fermion operator. One can check that this defines
$|\cdot,\cdot|$ unambiguously as a bilinear product on $\bar \lh$ and
that $\bar \lh$ is closed under the action of $|\cdot,\cdot|$. In
particular $|h_1,h_2|=[h_1,h_2]=0$ when any one of $h_1$ and $h_2$ is
a number. The set $\bar \lh$ equipped with the bilinear product
$|\cdot,\cdot|$ forms a so-called \emph{Lie superalgebra}. When only
fermions or only bosons are present, it becomes an ordinary Lie
algebra. The set
\begin{equation}
  \lh = \text{span} \{ \, ]a , b[ \; | \;
    a,b \in \aA \, \}
\end{equation}
can be shown to be a subalgebra of $\bar \lh$. Also the pointwise $G$
invariant subset of $\lh$, that is,
\begin{equation}
  \lh^G = \{ \, h \in \lh \; | \;
    \gamma h = h \gamma \quad \forall g \in G \, \}
\end{equation}
is a subalgebra because, as a subgroup of $\gGL(d)$, the group $G$
preserves the bracket $|\cdot,\cdot|$ on $\aA$. (The members of
$\gGL(d)$ just change the basis for the single-kind state space.)
Explicitly,
\begin{enumerate}

\item $\lh^{\gGL(d)}$ is spanned by the operators
\begin{multline}
   \sum_p \, ]a^\dagger_{p\tau},a_{p\upsilon}[ ,  \quad
    (\tau,\upsilon) \in K \times K \cup\bar K \times\bar K , \\
  \sum_p \, ]a^\dagger_{p\tau},a^\dagger_{p\upsilon}[ , \quad
  \sum_p \, ]a_{p\tau},a_{p\upsilon}[ , \quad
    (\tau,\upsilon) \in K \times\bar K ,
\end{multline}
where $K$ and $\bar K$ are the sets of $\tau$ with co- and
contragredient actions of $\gGL(d)$.

\item $\lh^{\gSp(d)}$ and $\lh^{\gO(d)}$ are spanned by the operators
\begin{equation}
  \sum_p \, ]a^\dagger_{p\tau},a_{p\upsilon}[ ,  \quad
  \sum_{pq} \, \langle p q | b \rangle \, 
    ]a^\dagger_{p\tau},a^\dagger_{q\upsilon}[ \, , \quad
  \sum_{pq} \, \langle b | p q \rangle \,
     ]a_{q\tau},a_{p\upsilon}[ .
\end{equation}

\end{enumerate}

Howe's theorem now reads:

\tit{General duality (Howe): The general Fock space has a
decomposition
$$
   \Phi = \bigoplus_\lambda \textup{X}_\lambda \otimes \Psi_\lambda ,
$$
where $G$ acts irreducibly on $\textup{X}_\lambda$ and $\lh^G$ acts
irreducibly on $\Psi_\lambda$. For $\lambda \ne \mu$, the
representations of $G$ on $\textup{X}_\lambda$ and $\textup{X}_\mu$
are inequivalent and the representations of $\lh^G$ on $\Psi_\lambda$
and $\Psi_\mu$ are inequivalent. The spaces $\textup{X}_\lambda$ have
finite dimensions.}

Howe's proof is based in the so-called \tit{first main
theorem}~\cite{ref:Wey39} of the classical groups, which states that
the algebra of their invariants is generated by the quadratic
invariants. Special cases include fermion and boson
$\gGL(d)$--$\lgl(k)$ dualities, fermion $\gSp(d)$--$\lsp(2k)$ and
$\gO(d)$--$\lo(2k)$ dualities and boson $\gSp(d)$--$\lo(2k)$ and
$\gO(d)$--$\lsp(2k)$ dualities. The boson $\lo(2k)$ and $\lsp(2k)$
irreps have infinite dimensions. In particular, the Lie algebra of the
$\gO(d)$--$\lsp(2k)$ duality is known in nuclear physics in the case
$d=A$ and $k=3$, where $A$ is the mass number, as the complexification
of the Lie algebra ``$\gSp(3,\mathbb R)$'' suggested by Rosensteel and
Rowe~\cite{ref:Ros77} to model nuclear collective motion. Since we are
in a boson environment, its irreps are infinite-dimensional. By Howe's
theorem, they can be labelled by the known equivalence classes of
finite-dimensional irreps of $\gO(A)$.

\section{Relation to the dualities with pairs of Lie
algebras\label{sec:rel}}

Unlike the dualities in Sections~\ref{sec:in} and \ref{sec:Hel}, which
relate a couple of Lie algebras, Howe duality relates a group and a
Lie (super-)algebra. Howe's theorem does not specify the relation of
the equivalence classes of the irreps carried by $\textup{X}_\lambda$
and $\Psi_\lambda$. This matter is addressed in later papers by
Howe~\cite{ref:How95}, Rowe, Repka and Carvalho~\cite{ref:Row11} and
me~\cite{ref:Nee20a}. It is fairly easy to show that in the special
cases with $\gGL(d)$ and $\gSp(d)$, the distinction between the group
and its Lie algebra does not matter. (The case of $\gGL(d)$ is the
most complicated one; see~\cite{ref:Nee20a}.) The case of $\gO(d)$ is
more involved due to the more complicated topology of this group. The
equivalence classes of $\gO(n)$ irreps were identified by
Weyl~\cite{ref:Wey39}. They are described by ordinary Young diagrams
subject to the constraint that no pair of different columns have
depths whose sum exceeds $n$. Rowe, Repka and Carvalho obtained the
relation between the equivalence classes of the $\gO(d)$ and $\lo(2k)$
irreps in the Howe $\gO(d)$--$\lo(2k)$ duality from an analysis of
highest weight states~\cite{ref:Row11}. In 2020, I derived it from the
$\lo(d)$--$\lo(2k)$ duality theorem in Section~\ref{sec:Hel} and
described it diagrammatically as follows~\cite{ref:Nee20a}.

\tit{$\gO(d)$--$\lo(2k)$ duality: A fermion Fock space has the
decomposition
\begin{equation}
   \Phi = \bigoplus \textup{X}_\lambda \otimes \Psi_\mu ,
\end{equation}
where $\gO(d)$ acts on $\textup{X}_\lambda$ producing an irrep with
the Young diagram $\lambda$ and $\lo(2k)$ acts on $\Psi_\mu$ producing
an irrep with the Young diagram $\mu$. The sum runs over the all such
pairs $(\lambda,\mu)$ that $\lambda$ and a rotated and reflected copy
of $\mu$ fill a $k \times d/2$ rectangle without overlap provided a
part of a row in $\mu$ of negative length cancels a part of $\lambda$
extruding the rectangle as in the following example for $d=11$ and
$k=4$, where $\lambda = (4,3,2,2,1,1,1)$ and
$\mu=(9/2,7/2,3/2,-3/2)$.}

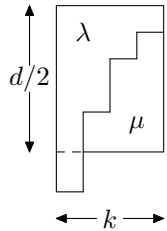
{\centering\includegraphics{young31}\par}

The symmetric way in which $\lo(d)$ and $\lo(2k)$ enter the
$\lo(d)$--$\lo(2k)$ duality theorem suggests that one might similarly
derive from this theorem a duality between $\lo(d)$ and a group. I
recently showed that this is indeed true and that the group is
$\gPin(2k)$~\cite{ref:Nee20b}. I first constructed a realisation of
$\gPin(2k)$ on the fermion Fock space from the observation that
$\gPin(2k)$ is realised within the Clifford algebra
$\aCl(k)$~\cite{ref:Bra35}, which is realised, in turn, by the algebra
generated by the operators $a^\dagger_{1\tau}$ and $a_{1\tau}$. More
precisely, $\gPin(2k)$ can be identified with the set of products of
linear combinations $s = \sum_\tau ( \alpha_\tau a^\dagger_{1\tau}
+ \beta_\tau a_{1\tau})$ obeying $s^2=-1$ and $\gSpin(2k)$ with the
set of products of an even number of factors of this
form~\cite{ref:Goo98}. A single $s$ thus connects the subgroup
$\gSpin(2k)$ and its coset. In particular, in my realisation on the
fermion Fock space, $s = a^\dagger_{11} - a_{11}$ maps to a ``partial
particle-hole conjugation'' $\sigma$ obeying
\begin{multline}\label{eq:siga}
  \sigma a_{p1} = (-)^d a^\dagger_{p1} \sigma , \quad
  \sigma a^\dagger_{p1} = (-)^d a_{p1} \sigma , \\
  \sigma a_{p\tau} = (-)^d a_{p\tau} \sigma , \quad
  \sigma a^\dagger_{p\tau} = (-)^d a^\dagger_{p\tau} \sigma , \quad
    \tau > 1 .
\end{multline}

One can \emph{define} a set of generalised Young diagrams by the
following rules. (i)~The rows have either integral or half-integral,
positive row lengths, which decrease weakly from top to bottom.
(ii)~If the row lengths are integral, no pair of different columns
have depths whose sum exceeds $2k$. (iii)~If the row lengths are
half-integral, the Young diagram has exactly $k$ rows. One can assign
to each such Young diagram properties of a $\gPin(2k)$ irrep, which my
limited space does not allow me to specify. (See
ref.~\cite{ref:Nee20b} for details.) Essentially, the Young diagrams
with integral row lengths describe $\gPin(2k)$ irreps which factor
through $\gO(2k)$, yielding an $\gO(2k)$ irrep with the same diagram,
while those with half-integral row lengths describe $\gPin(2k)$ irreps
which split upon restriction into a pair of mirror $\lo(2k)$ irreps,
one of which has the same Young diagram. My constructed realisation of
$\gPin(2k)$ on the fermion Fock space being denoted by the same
symbol, I then proved the following.

\tit{$\lo(d)$--$\gPin(2k)$ duality: A fermion Fock space has the
decomposition
$$
   \Phi = \bigoplus \textup{X}_\lambda \otimes \Psi_\mu ,
$$
where $\lo(d)$ acts on $\textup{X}_\lambda$ producing an irrep with
the Young diagram $\lambda$ and $\gPin(2k)$ acts on $\Psi_\mu$
producing an irrep with the Young diagram $\mu$. The sum runs over all
such pairs $(\lambda,\mu)$ that $\lambda$ and a rotated and reflected
copy of $\mu$ fill a $k \times d/2$ rectangle without overlap provided
a part of a row in $\lambda$ of negative length cancels a part of
$\mu$ extruding the rectangle as in the following example for $d=12$
and $k=4$, where $\lambda = (4,3,2,2,1,-1)$ and $\mu=(5,4,2,1,1)$.}

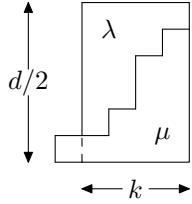
{\centering\includegraphics{young32}\par}

\section{Concluding remarks\label{sec:con}}

The triple of the $\lo(d)$--$\lo(2k)$, $\gO(d)$--$\lo(2k)$ and
$\lo(d)$--$\gPin(2k)$ dualities seem to present a nice, unified
picture with a high degree of symmetry between the number conserving
and the number non-conserving groups or Lie algebras. The only
asymmetry is $\gO(d)$ versus $\gPin(2k)$, or equivalently, the
presence of only non-spin irreps of $\lo(d)$. This asymmetry is
related to $2k$ being always even while both even and odd values are
allowed for $d$. My proofs of the $\gO(d)$--$\lo(2k)$ and
$\lo(d)$--$\gPin(2k)$ dualities is based on the $\lo(d)$--$\lo(2k)$
duality, whose proof is based, in turn, on a comparison of characters.
In particular the proof of the $\lo(d)$--$\gPin(2k)$ duality does not
require a $\gPin(n)$ first main theorem, and in fact, such a theorem
is not known~\cite{ ref:How94}. Application of the method is
restricted, though, to the fermion case because it employs Weyl's
formula~\cite{ref:Jac62} for the character of a finite-dimensional
irrep of a semi-simple Lie algebra over the complex numbers.

\end{document}